\documentclass[aps,pre,amsmath,twocolumn,superscriptaddress,showpacs]{revtex4-1}
\usepackage{fancyhdr}
\usepackage{calc}
\usepackage{amsmath}
\usepackage{amsfonts}
\usepackage{amssymb}
\usepackage{graphicx}
\usepackage{hyperref}
\usepackage{soul}
\usepackage{mathtools}
\usepackage{verbatim}
\usepackage{epstopdf}
\usepackage{subfigure}
\usepackage{latexsym}
\usepackage{dcolumn}
\usepackage{epsf}
\usepackage{float}
\usepackage[table]{xcolor}
\usepackage{multirow}

\DeclarePairedDelimiter\norm{\lVert}{\rVert}
\usepackage{color} 

\usepackage{graphicx}
\usepackage{graphicx,epstopdf}
\usepackage{xcolor}
\usepackage{dcolumn}
\usepackage{bm}
\usepackage{mathrsfs} 

\usepackage{mathrsfs}
\begin{document}
\title{Interlayer antisynchronization in degree-biased duplex networks}

\author{Sayantan Nag Chowdhury}
\affiliation{Department of Environmental Science and Policy, University of California, Davis, California 95616, USA}
\affiliation{Technology Innovation Hub (TIH), IDEAS (Institute of Data Engineering Analytics and Science Foundation), @ Indian Statistical Institute, 203 B. T. Road, Kolkata-700108, India}
\email{jcjeetchowdhury1@gmail.com}
\author{Sarbendu Rakshit}
\affiliation{Department of Mechanical Engineering, University of California, Riverside, California 92521, USA}
\author{Chittaranjan Hens}
\affiliation{Center for Computational Natural Sciences and Bioinformatics, International Institute of Information Technology, Gachibowli, Hyderabad-500032, India}
\email{chittaranjanhens@gmail.com}
\author{Dibakar Ghosh}
\affiliation{Physics and Applied Mathematics Unit, Indian Statistical Institute, 203 B. T. Road, Kolkata-700108, India}
\email{dibakar@isical.ac.in}


	

\date{\today}
\begin{abstract}

	With synchronization being one of nature's most ubiquitous collective behaviors, the field of network synchronization has experienced tremendous growth, leading to significant theoretical developments. However, most of these previous studies consider uniform connection weights and undirected networks with positive coupling. In the present article, we incorporate the asymmetry in a two-layer multiplex network by assigning the ratio of the adjacent nodes' degrees as the weights to the intralayer edges. Despite the presence of degree-biased weighting mechanism and attractive-repulsive coupling strengths, we are able to find the necessary conditions for intralayer synchronization and interlayer antisynchronization and test whether these two macroscopic states can withstand demultiplexing in a network. During the occurrence of these two states, we analytically calculate the oscillator's amplitude. {\color{black}In addition to deriving the local stability conditions for interlayer antisynchronization via the master stability function approach, we also construct a suitable Lyapunov function to determine a sufficient condition for global stability.} 
	 We provide numerical evidence to show the necessity of negative interlayer coupling strength for the occurrence of antisynchronization, and such repulsive interlayer coupling coefficients can not destroy intralayer synchronization.

\end{abstract}


\maketitle 

\section{Introduction}

\par Multilayer networks \cite{kivela2014multilayer,boccaletti2014structure,wang2015evolutionary} of coupled oscillators provide a fascinating platform to study the collective asymptotic behavior of dynamical systems evolving on top of it. Several layers of such a network prove to be a fertile playground to reveal the interplay between the network structure and the unfolding of collective phenomena of various dynamical processes. 
The hallmark property of a realistic system is the complex connectivity patterns of its components, and it may often give rise to complex dynamics. An isolated network can seldom describe such collective dynamics of interconnected systems. Thus, researchers often resort to multilayer networks anticipating some new fresh insights into complex systems. In the past years, numerous studies have unfolded several emergent collective phenomena, such as extreme events \cite{chen2015extreme,hernandez2013probabilistic}, percolation \cite{gao2012networks,buldyrev2010catastrophic}, congestion of traffic \cite{helbing2001traffic,morris2012transport}, epidemics spreading \cite{granell2013dynamical,sanz2014dynamics}, controllability \cite{menichetti2016control}, evolutionary game dynamics \cite{gomez2012evolution,nag2020cooperation}, and diffusion \cite{gomez2013diffusion}, to name a few. The results presented in these studies demonstrate a very different phenomenology from the one found in monolayer networks. Various complex forms of synchronized dynamics of multilayer networks of the coupled oscillator, as for instance interlayer synchronization \cite{leyva2017inter,rakshit2020invariance}, relay synchronization \cite{leyva2018relay}, antiphase synchronization \cite{chowdhury2021antiphase}, relay interlayer synchronization \cite{rakshit2021relay},  intralayer synchronization \cite{gambuzza2015intra,rakshit2020intralayer,anwar2022intralayer}, cluster synchronization \cite{della2020symmetries,jalan2016cluster}, explosive synchronization \cite{zhang2015explosive,jalan2019explosive,khanra2018explosive}, breathing synchronization \cite{louzada2013breathing}, solitary states \cite{majhi2019solitary}, and complete synchronization \cite{del2016synchronization}, have been brought to the limelight by investigating the role that network structure plays in the onset and stability of such coherent states. Nevertheless, the study of interlayer antisynchronization on multilayer structures remains relatively unexplored to the best of our knowledge under different contexts.

\par Interlayer antisynchronization in a multiplex network refers to the dynamical process where two identical oscillators directly connected through the interlayer link settle down to an equal amplitude with a constant phase difference of $\pi$. 
 Inspired by antiphase patterns in two-module neuronal networks \cite{li2011organization}, we are interested in deriving the criteria for the existence and stability of interlayer antisynchronization state in a duplex (multiplex with two layers). {\color{black}Apart from performing local stability analysis of this state of the interacting systems with the help of the master stability function (MSF) approach \cite{pecora1998master,tang2019master}, we are equally interested in deriving the sufficient condition for global stability of interlayer antisynchronization state. 
 To do this, we construct a suitable Lyapunov function for deriving the global stability of this state.} The phrase `global stability' here reflects that the system will evolve into the interlayer antisynchronization state irrespective of the chosen initial conditions except for a set of measure zero \cite{kassabov2022global}.

\par Most of the previous investigations on the synchronization \cite{chowdhury2019synchronization2, arenas2008synchronization,chowdhury2019convergence,wu2022double,chowdhury2021extreme,ghosh2022synchronized} of complex networks of coupled dynamical systems are performed by assuming (i) unweighted and undirected networks and (ii) attractive (positive) coupling strengths only. However, realistic systems are far more complicated, and there are ample real-life examples where heterogeneous connectivity weights \cite{chatterjee2022controlling} and the simultaneous presence of attractive-repulsive interactions \cite{mishra2015chimeralike,hens2013oscillation,hens2014diverse,bera2016emergence,kundu2019emergent} are beneficial in portraying real-world scenarios. For instance, the number of emails exchanged between two colleagues in an organization, and the number of scientific collaborations between two scientists depend on different contexts. It is best to assign a weight to each edge of the network to derive relationships between such interacting individuals. Instead of using the random weighted directed network, we consider the influence of a node's degree on its neighbors and construct a degree-biased network to study the interlayer antisynchronization and intralayer synchronization in the multiplex. Synchronization on weighted networks have been studied extensively in the literature, as indicated in the following references \cite{zhou2006universality,zhou2006dynamical,chavez2006synchronizing,kempton2017self,leyva2013explosive}. Nevertheless, to our knowledge, the emergence and (local and global) stability of interlayer antisynchronization on multiplex networks with weighted intralayer connections have never been investigated.   Furthermore, we introduce the negative interlayer coupling strength, which is found to be essential for the onset of the interlayer antisynchronization as per our numerical simulations. The positive intralayer coupling strengths allow the system to settle into the intralayer synchronization, despite the presence of negative interlayer coupling strength. Numerous real-life scenarios are highlighted in the review \cite{majhi2020perspective} to emphasize the importance of attractive-repulsive interaction. As per Ref.\ \cite{chowdhury2020effect}, all the pairs of interacting subunits of a system can not minimize their energy due to opposing coupling strengths. When network connections change over time, such temporal networks with positive-negative coupling may produce several peculiar states like static $\pi$ state \cite{sar2022swarmalators}, extreme events \cite{chowdhury2019synchronization, chowdhury2022extreme}, inhomogeneous small oscillation \cite{chowdhury2020distance}, and many more. Ecologists and data analysts also unveil the tug of war between positive and negative interactions for extracting useful information about ecosystems' diversity in species \cite{giron2016synchronization,bacelar2014exploring}.

\par Following the seminal works by Estrada and his collaborators \cite{estrada2020hubs,gambuzza2020hubs}, we consider three distinct types of intralayer networks, viz.\ hubs-attracting, hubs-repelling, and unweighted network. We furnish analytical insights about the conditions for the emergence of intralayer synchronization and interlayer antisynchronization. We analytically derive the necessary conditions for all the identical oscillators to evolve in unison within the layers. All these analytical results help to design a duplex with suitable oscillators and couplings that allows the system to achieve such coherent states. Our numerical simulations also support that our analytical findings (existence and stability criteria) effectively help to achieve intralayer synchronization and interlayer antisynchronization when appropriate conditions are met.

\section{Mathematical model} \label{model}

\par To illustrate our findings, we consider a multiplex network with two layers. On top of the vertices of each layer consisting of $N$ nodes, we place an $m$-dimensional identical dynamical system with state vectors ${\bf x}_{\alpha,i} \in \mathbb{R}^m$, $\alpha=1,2$ and $i=1,2,3,\cdots,N$. Here, the first component (i.e., $\alpha$) of the suffices of ${\bf x}_{\alpha,i}$ represents the number of the layer, and the second component (i.e., $i$) depicts the number of the node of the $\alpha$-th layer. Each of these isolated oscillators maintains the dynamical equations in the absence of intralayer and interlayer couplings as follows

\begin{equation} \label{1}
\begin{split}
\dot{\bf x}_{\alpha,i}=f({\bf x}_{\alpha,i}),
\end{split}
\end{equation}
where $f:\mathbb{R}^m\to\mathbb{R}^m$ is the autonomous nonlinear evolution function. We assume this $f$ is continuously differentiable with respect to its argument. We need to consider this assumption, which we need later for performing the stability analysis. Let $\mathscr{A}_{ij}^{[\alpha]}$, $\alpha=1,2$ be the elements of the adjacency matrix encoding the intralayer topology of the $\alpha$-th layer. {\color{black}Precisely, for $\alpha=1,2$; 

\begin{equation}
	\mathscr{A}_{ij}^{[\alpha]}=
	\begin{cases}
		1, & \text{if} \hspace*{0.1cm}i \hspace*{0.1cm} \text{-th} \hspace*{0.1cm} \text{and}\hspace*{0.1cm} j \text{-th nodes are connected in the}\hspace*{0.1cm} \alpha \text{-th layer} \\
		0, & \text{otherwise } 
	\end{cases}
\end{equation}

} Since we are also interested in inspecting intralayer synchronization, we only consider connected intralayer networks. When both the layers are coupled, then we can describe the dynamical evolution of the $i$-th node of $\alpha$-th layer as follows,

\begin{equation} \label{2}
\begin{split}
\dot{\bf x}_{1,i}=f({\bf x}_{1,i})+k_A\sum_{j=1}^{N}\mathscr{\tilde{A}}_{ij}^{[1]} G[{\bf x}_{1,j},{\bf x}_{1,i}]+k_R H[{\bf x}_{2,i},{\bf x}_{1,i}],\\
\dot{\bf x}_{2,i}=f({\bf x}_{2,i})+k_A\sum_{j=1}^{N}\mathscr{\tilde{A}}_{ij}^{[2]} G[{\bf x}_{2,j},{\bf x}_{2,i}]+k_R H[{\bf x}_{1,i},{\bf x}_{2,i}].
\end{split}
\end{equation}
Here, $\mathscr{\tilde{A}}_{ij}^{[\alpha]}$ is generated by assigning a weight to each element $\mathscr{A}_{ij}^{[\alpha]}$ as follows

\begin{equation} \label{3}
\begin{split}
\mathscr{\tilde{A}}_{ij}^{[\alpha]}={\bigg(\dfrac{d_j}{d_i}}\bigg)^{\beta} \mathscr{A}_{ij}^{[\alpha]},
\end{split}
\end{equation}
where $\beta \in \{0,1,-1\}$ and $d_i$ denotes the degree of the $i$-th node in the whole multiplex network whose adjacency matrix is given by

\begin{equation} \label{4}
\mathscr{A}=\begin{pmatrix}
\mathscr{A}^{[1]} & I \\ 
I & \mathscr{A}^{[2]} \\	
\end{pmatrix}.
\end{equation} 

Here, $I$ is the identity matrix of order $N$. When $\beta=+1$, we have a hub-attracting intralayer adjacency matrix by adopting the terminology from Ref.\ \cite{gambuzza2020hubs}. This rescaled unsymmetric hub-attracting matrix reflects the tendency to produce a strong influence on the low-degree neighbors by the high-degree nodes \cite{anwar2021relay}. We can inspect the reverse scenario of biased domination from low to high-degree nodes with the hub-repelling matrix by considering $\beta=-1$ \cite{anwar2021relay,estrada2020hubs}. However, the matrix remains unaltered for $\beta=0$, i.e., we have $\mathscr{\tilde{A}}_{ij}^{[\alpha]}=\mathscr{{A}}_{ij}^{[\alpha]}$ for $\beta=0$. Thus, Eq.\ \ref{2} reduces to

\begin{equation} \label{5}
\begin{split}
\dot{\bf x}_{1,i}=f({\bf x}_{1,i})+k_A\sum_{j=1}^{N}{\bigg(\dfrac{d_j}{d_i}}\bigg)^{\beta} \mathscr{{A}}_{ij}^{[1]} G[{\bf x}_{1,j},{\bf x}_{1,i}]\\+k_R H[{\bf x}_{2,i},{\bf x}_{1,i}],\\
\dot{\bf x}_{2,i}=f({\bf x}_{2,i})+k_A\sum_{j=1}^{N}{\bigg(\dfrac{d_j}{d_i}}\bigg)^{\beta} \mathscr{{A}}_{ij}^{[2]} G[{\bf x}_{2,j},{\bf x}_{2,i}]\\+k_R H[{\bf x}_{1,i},{\bf x}_{2,i}].
\end{split}
\end{equation}

Here, $k_A$ is the intralayer coupling strength, and $G:\mathbb{R}^m\times\mathbb{R}^m\to\mathbb{R}^m$ is the output vectorial function within the layers. On the other hand, $k_R$ is the interlayer coupling strength, and $H:\mathbb{R}^m\times\mathbb{R}^m\to\mathbb{R}^m$ is the interlayer coupling vectorial function. In the next section, we rigorously investigate the necessary criteria for the existence of interlayer antisynchronization state and intralayer synchronization.

\section{Analytical findings}\label{Analytical}

\par Before representing our key analytical findings, first, we briefly define two synchronized states, viz.\ (i) interlayer antisynchronization and (ii) intralayer synchronization states.

The interlayer antisynchronization depicts the synchronous time evolution of the oscillators situated on top of the replica nodes with a constant phase difference of $\pi$, i.e., the sum $({\bf x}_{1,i}(t)+{\bf x}_{2,i}(t))$ of the dynamics of the state variables of the $i$-th oscillators of both layers vanishes after the transient. Mathematically, when the system evolves in the interlayer antisynchronization state, we have  

\begin{equation} \label{6}
{\bf x}_{1,i}(t)+{\bf x}_{2,i}(t)={\bf 0}, \hspace{0.5cm} \forall \hspace{0.3cm} i=1,2,\cdots,N.
\end{equation}

We define the interlayer antisynchronization error as the following

\begin{equation}\label{7}
E=\lim\limits_{t\to\infty}\dfrac{1}{T}\int_{t}^{t+T}\sum\limits_{i=1}^{N}\dfrac{\norm{{\bf x}_{1,i}(t)+{\bf x}_{2,i}(t)}}{N}~d\tau
\end{equation}

Clearly, $E$ necessarily becomes zero in the state of interlayer antisynchronization and remains non-zero otherwise.

\par On the other hand, intralayer synchronization remains completely independent of the interlayer antisynchronization. A system may evolve in the interlayer antisynchronization state; however, it may not maintain the intralayer synchronization and vice-versa. Intralayer synchronization refers to the synchronous evolution of all dynamical units within each layer. In other words, proceeding to the limit as $t\to\infty$ for all $i=1,2,\cdots,N$ and $\alpha=1,2$, there definitely exists intralayer synchronization solution ${\bf x}_{\alpha}(t)\in\mathbb{R}^m$ 
such that

\begin{equation} \label{8}
{\bf x}_{\alpha,i}(t) \to {\bf x}_{\alpha}(t)
\end{equation}

Now, we move on to prove the necessary conditions on the individual node dynamics $f$, the network topology $\mathscr{\tilde{A}}_{ij}^{[\alpha]}$, the coupling functions $G$ and $H$ for the emergence of interlayer antisynchronization and intralayer synchronization states. 

\subsection{Necessary condition for interlayer anti synchronization state}\label{Interlayer synchronization}

\par When the system evolves in the interlayer antisynchronization state , all the vertices of one layer maintain the same amplitude with its replica nodes of the different layers. Still, their phase difference will be $\pi$. Thus, using Eq.\ \eqref{6}, we obtain the following set of equations from Eq.\ \eqref{2}  as follows

\begin{equation} \label{9}
\begin{split}
\dot{\bf x}_{1,i}=f({\bf x}_{1,i})+k_A\sum_{j=1}^{N}\mathscr{\tilde{A}}_{ij}^{[1]} G[{\bf x}_{1,j},{\bf x}_{1,i}]\\+k_R H[-{\bf x}_{1,i},{\bf x}_{1,i}],\\
\dot{\bf x}_{1,i}=-f(-{\bf x}_{1,i})-k_A\sum_{j=1}^{N}\mathscr{\tilde{A}}_{ij}^{[2]} G[-{\bf x}_{1,j},-{\bf x}_{1,i}]\\-k_R H[{\bf x}_{1,i},-{\bf x}_{1,i}].
\end{split}
\end{equation}

These equations remain consistent if
\begin{enumerate}
	\item $f(\bf{x})$$=-f(-\bf{x})$, i.e., $f$ is an odd function,
	\item $\sum_{j=1}^{N}\mathscr{\tilde{A}}_{ij}^{[1]} G[{\bf x}_{1,j},{\bf x}_{1,i}]\\=-\sum_{j=1}^{N}\mathscr{\tilde{A}}_{ij}^{[2]} G[-{\bf x}_{1,j},-{\bf x}_{1,i}]$, and
	\item $H(-\bf{x},\bf{x})$$=-H(\bf{x},-\bf{x})$, i.e., $H$ is an odd function.
\end{enumerate}

These three conditions are necessary for obtaining interlayer antisynchronization and, by no means, sufficient ones. Mere fulfilling these three conditions, one can not anticipate interlayer antisynchronization.

\subsection{Necessary condition for intralayer synchronization}\label{Intralayer}

\par Let all the trajectories of the first layer maintain a coherent rhythm, i.e., ${\bf x}_{1,i}(t)$ converges to ${\bf x}_{1}(t)$ at some time $t=t_{1}$ (say). Similarly, ${\bf x}_{2,i}(t)$ of the second layer converges to ${\bf x}_{2}(t)$ at some time $t=t_{2}$. Let $t_{0}$ be the maximum of $\{t_{1},t_{2}\}$. Thus, for any time $t \geq t_{0}$, the rate of changes of all the state variables in all respective layers should be identical. The system converges into the intralayer synchronization manifold $({\bf x}_{1}(t),{\bf x}_{2}(t))$ for $t \geq t_{0}$. 

\par Without loss of any generality, we choose two arbitrary nodes $i$ and $l$ (say) from both the layers. Therefore, we have ${\bf x}_{1,i}(t)={\bf x}_{1,l}(t)={\bf x}_{1}(t)$ and ${\bf x}_{2,i}(t)={\bf x}_{2,l}(t)={\bf x}_{2}(t)$, once the system \eqref{2} settles into the intralayer synchronization manifold. Then the corresponding dynamics of the $i$-th and $l$-th nodes of the first layer are governed by the following ordinary differential equations

\begin{equation} \label{10}
\begin{split}
\dot{\bf x}_{1}=\dot{\bf x}_{1,i}=f({\bf x}_{1})+k_A\sum_{j=1}^{N}\mathscr{\tilde{A}}_{ij}^{[1]} G[{\bf x}_{1},{\bf x}_{1}]\\+k_R H[{\bf x}_{2},{\bf x}_{1}],\\
\dot{\bf x}_{1}=\dot{\bf x}_{1,l}=f({\bf x}_{1})+k_A\sum_{j=1}^{N}\mathscr{\tilde{A}}_{lj}^{[1]} G[{\bf x}_{1},{\bf x}_{1}]\\+k_R H[{\bf x}_{2},{\bf x}_{1}].
\end{split}
\end{equation}

Substracting these two equations, we obtain

\begin{equation} \label{11}
\begin{split}
\sum_{j=1}^{N}(\mathscr{\tilde{A}}_{ij}^{[1]}-\mathscr{\tilde{A}}_{lj}^{[1]}) G[{\bf x}_{1},{\bf x}_{1}]=0.
\end{split}
\end{equation}

Similarly, the dynamics of the $i$-th and $l$-th nodes of the second layer yield the following equation

\begin{equation} \label{12}
\begin{split}
\sum_{j=1}^{N}(\mathscr{\tilde{A}}_{ij}^{[2]}-\mathscr{\tilde{A}}_{lj}^{[2]}) G[{\bf x}_{2},{\bf x}_{2}]=0.
\end{split}
\end{equation}

Since both the two chosen nodes $i$ and $l$ are arbitrary, thus the necessary condition for the intralayer synchronization gives the following criteria

$\sum_{j=1}^{N}\mathscr{\tilde{A}}_{ij}^{[\alpha]}=\sum_{j=1}^{N}\mathscr{\tilde{A}}_{lj}^{[\alpha]}$, $\alpha=1,2$, i.e., the in-degree of each node in the each layer must be equal.\\
or, $G[{\bf x}_{\alpha},{\bf x}_{\alpha}]=0$, $\alpha=1,2$, i.e., the intralayer coupling function $G$ vanishes after the oscillators of each layer evolve synchronously.


\section{Results} \label{numerical}

\par For numerical simulations, we utilize FORTRAN 90 compiler. We integrate Eq.\ \eqref{2} using the fifth-order Runge-Kutta-Fehlberg method with integration time step $h = 0.01$. As per our derived necessary conditions on the interlayer antisynchronization, $H$ needs to be an odd function. Hence, we choose $H({\bf x}_i,{\bf x}_j)=[{x}_j + {x}_i, {y}_j + {y}_i]^T$ 
where $T$ represents the transpose of a vector. {\color{black}Similarly, the necessary conditions for the intralayer synchronization reveal either the in-degree of each node of the intralayer network is equal or $G$ should vanish after the intralayer synchronization is achieved. Hence, we choose $G({\bf x}_i,{\bf x}_j)=[{x}_j - {x}_i, {y}_j - {y}_i]^T$ as in the form of the linear diffusive coupling, so that $G$ will become identically zero after achieving the intralayer synchronization state. The diffusive coupling was previously used in many systems \cite{pikovsky2001universal}, which removes the restriction on the intralayer network connectivity. This choice of $G$ will allow us to choose any connected intralayer network.}

\subsection{Stuart-Landau Oscillators}

\par We first choose identical Stuart-Landau (SL) oscillators \cite{kuramoto2003chemical} to begin our numerical investigations. The state dynamics of the limit cycle oscillator situated on top of the $i$-th node is represented by 
\begin{equation} \label{13}
f({\bf x}_i)=\left(
\begin{array}{c}
\left[1-\left({x_i}^2+{y_i}^2\right)\right]x_i-\omega_i y_i\\\\
\left[1-\left({x_i}^2+{y_i}^2\right)\right]y_i+\omega_i x_i\\
\end{array}
\right), \\
\end{equation}

where ${\bf x}_i \in \mathbb{R}^{2}$. Since we are basically interested in the interlayer antisynchronization and intralayer synchronization, thus we choose the same intrinsic frequency $\omega_i=\omega=3$ for all oscillators. 
Clearly, this $f$, being the odd function, satisfies the necessary condition for the emergence of the interlayer antisynchronization state.

\subsubsection{Amplitude of each oscillator maintaining interlayer antisynchronization and intralayer synchronization}

\par We analytically calculate the amplitude of each SL oscillator when each oscillator in a single layer undergoes a synchronous evolution with all the other units of the same layer, and simultaneously, each oscillator maintains an antisynchronization state with all its replicas in different layers. The chosen functions $G$ and $H$ help us to write the dynamical evolution of each $l$-th SL oscillator $(l=1,2,\cdots,N)$ in the $\alpha$-th layer $(\alpha=1,2)$ in terms of the complex variable $z_{\alpha,l}=x_{\alpha,l}+ky_{\alpha,l}=r_{\alpha,l} e^{k{\theta}_{\alpha,l}}  \in \mathbb{C}$ as follows

\begin{equation} \label{14}
\begin{array}{lll}
\dot{z}_{1,l}=&(1-|z_{1,l}|^2)z_{1,l}+k\omega z_{1,l}+k_A\sum_{j=1}^{N}\mathscr{\tilde{A}}_{lj}^{[1]}(z_{1,j}-z_{1,l})\\[2pt]&+k_R (z_{2,l}+z_{1,l}),\\[5pt]
\dot{z}_{2,l}=&(1-|z_{2,l}|^2)z_{2,l}+k\omega z_{2,l}+k_A\sum_{j=1}^{N}\mathscr{\tilde{A}}_{lj}^{[2]}(z_{2,j}-z_{2,l})\\[2pt]&+k_R (z_{1,l}+z_{2,l}),
\end{array}
\end{equation}

where $k=\sqrt{-1}$, $r_{\alpha,l}=\sqrt{{x}^2_{\alpha,l}+{y}^2_{\alpha,l}}$ is the amplitude of the SL oscillator situated in the $l$-th node of the $\alpha$-th layer and the phase of that SL oscillator, ${\theta}_{\alpha,l}$ is given by the principal value of argument of the complex number $z_{\alpha,l}$, i.e., ${\theta}_{\alpha,l}=\tan^{-1}\bigg(\dfrac{y_{\alpha,l}}{x_{\alpha,l}}\bigg)$. By substituting $z_{1,l}=r_{1,l} e^{k{\theta}_{1,l}}$ and $z_{2,l}=r_{2,l} e^{k{\theta}_{2,l}}$ in \eqref{14}, we find the phase of the oscillators obeys the following ordinary differential equations

\begin{equation}\label{15}
\begin{split}
\dot{{\theta}}_{1,l}=\omega+k_A\sum_{j=1}^{N}\mathscr{\tilde{A}}_{lj}^{[1]}\frac{r_{1,j}}{r_{1,l}}\sin{({\theta}_{1,j}-{\theta}_{1,l})}\\+k_R\frac{r_{2,l}}{r_{1,l}}\sin{({\theta}_{2,l}-{\theta}_{1,l})},\\
\dot{{\theta}}_{2,l}=\omega+k_A\sum_{j=1}^{N}\mathscr{\tilde{A}}_{lj}^{[2]}\frac{r_{2,j}}{r_{2,l}}\sin{({\theta}_{2,j}-{\theta}_{2,l})}\\+k_R \frac{r_{1,l}}{r_{2,l}}\sin{({\theta}_{1,l}-{\theta}_{2,l})}.
\end{split}
\end{equation}

In order to obtain these equations, we assume $r_{\alpha,l} \neq 0$, $l=1,2,\cdots,N$ and $\alpha=1,2$. Clearly, if $r_{\alpha,l}=0$, then the system converges to the origin giving rise to the amplitude death state \cite{saxena2012amplitude,dixit2021emergent,resmi2011general,dixit2021dynamic}. Hence for $r_{\alpha,l}=0$, we can not anticipate interlayer antisynchronization state. Thus, we neglect the case of $r_{\alpha,l}=0$. 
Similarly, we derive the rate of change of amplitude of the $l$-th SL oscillator as follows,

\begin{equation}\label{16}
\begin{split}
\dot{r}_{1,l}=(1-{r}^2_{1,l})r_{1,l}+k_A\sum_{j=1}^{N}\mathscr{\tilde{A}}_{lj}^{[1]}(r_{1,j} \cos{({\theta}_{1,j}-{\theta}_{1,l})}-r_{1,l})\\+k_R(r_{2,l}\cos{({\theta}_{2,l}-{\theta}_{1,l})}+r_{1,l}),\\
\dot{r}_{2,l}=(1-{r}^2_{2,l})r_{2,l}+k_A\sum_{j=1}^{N}\mathscr{\tilde{A}}_{lj}^{[2]}(r_{2,j} \cos{({\theta}_{2,j}-{\theta}_{2,l})}-r_{2,l})\\+k_R(r_{1,l}\cos{({\theta}_{1,l}-{\theta}_{2,l})}+r_{2,l}).
\end{split}
\end{equation}

\begin{figure}[ht]
	\centerline{\includegraphics[scale=0.3]{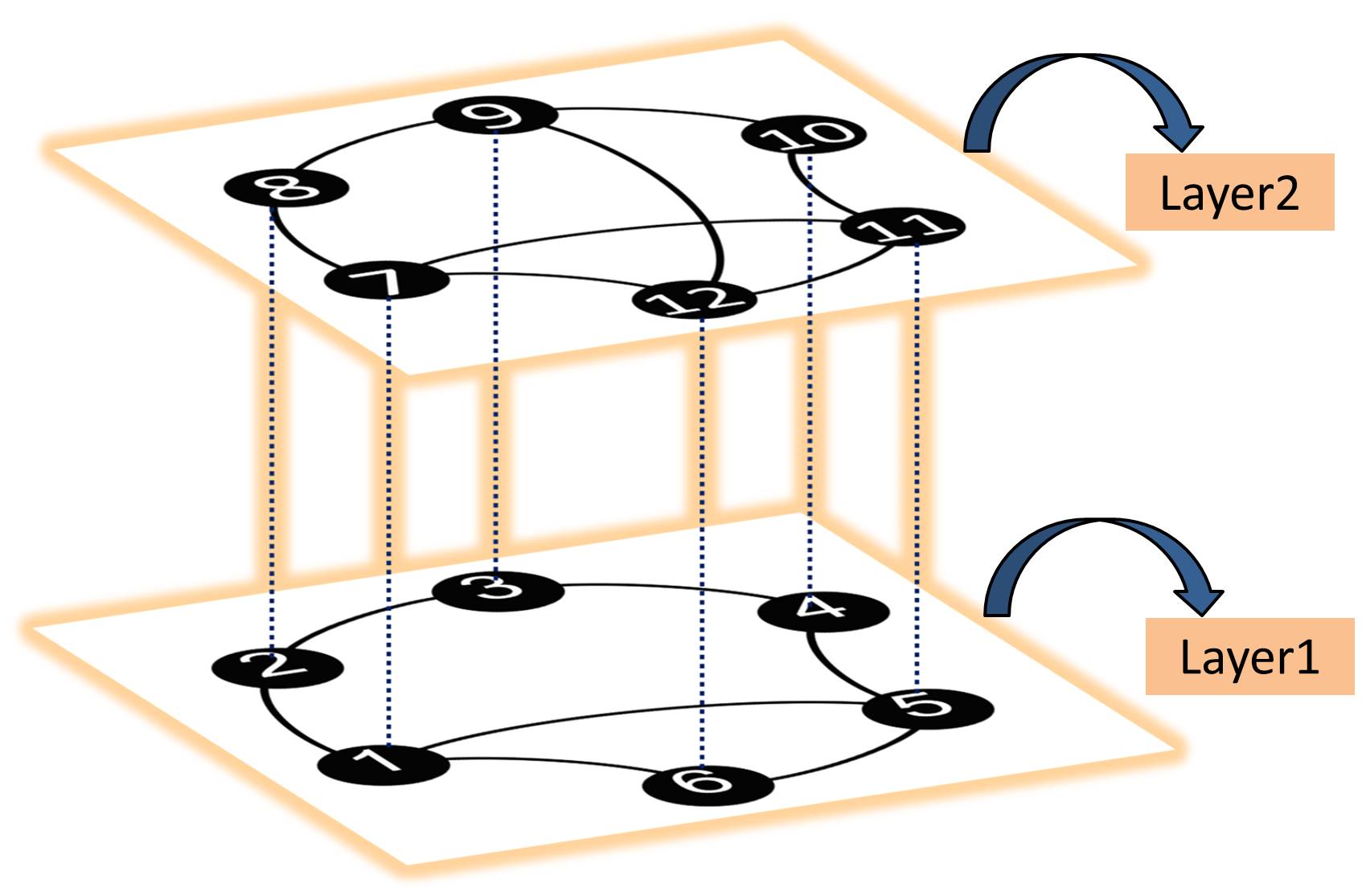}}
	\caption{{\bf A multiplex network}: We here visualize a duplex network (a multiplex network with two layers) with the help of Gephi \cite{bastian2009gephi}. Each of these layers consists of a connected intralayer network. This network used for the numerical experiments (unless stated otherwise) contains $12$ nodes and $21$ links. The six interlayer edges (dotted lines) connect the replica nodes and help to connect the two connected layers.}\label{fig1}
\end{figure}

For complete intralayer synchronization state, we have

\begin{equation}\label{17}
\begin{split}
r_{\alpha,i}=r_{\alpha},\\
\theta_{\alpha,i}=\theta_{\alpha}
\end{split}
\end{equation}

for $\alpha=1,2$ and $i=1,2,\cdots,N$.

Furthermore, if the system evolves in the interlayer antisynchronization state, then we have

\begin{equation}\label{18}
\begin{split}
r_1=r_2,\\
{\theta}_{1}-{\theta}_{2}= \pm \pi.
\end{split}
\end{equation}

Using Eqs.\ \eqref{17} and \eqref{18}, Eq.\ \eqref{16} becomes

\begin{equation}\label{19}
\begin{split}
\dot{r}_1=(1-{r}^{2}_{1})r_1.
\end{split}
\end{equation}

Thus, the duplex networks in the presence of intralayer synchronization and interlayer antisynchronization states can be described by Eq.\ \eqref{19}, where the local dynamics of each node are associated with the SL oscillator \eqref{13}. Solving Eq.\ \eqref{19} as a function of time $t$, we have

\begin{equation}\label{20}
\begin{split}
{r}_1=+\sqrt{\dfrac{e^{2t}}{e^{2t}-c_1}}.
\end{split}
\end{equation}

\begin{figure*}[ht]
	\centerline{\includegraphics[scale=0.5]{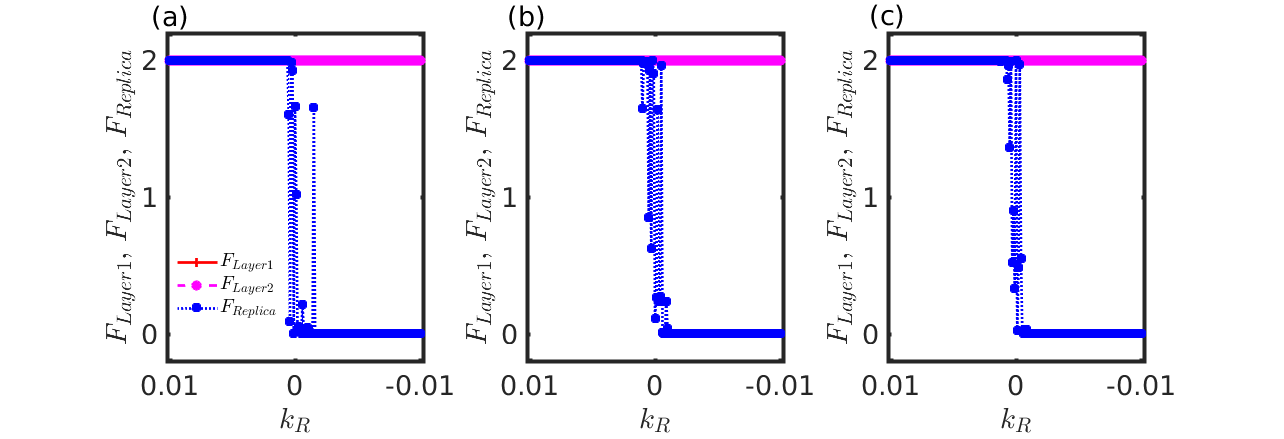}}
	\caption{{\bf The variation of $F_{Layer1}$, $F_{Layer2}$, and $F_{Replica}$ as a function of interlayer coupling strength $k_R$}: We choose the multiplex network shown in Fig.\ \eqref{fig1}, and place an identical limit cycle oscillator \eqref{13} on top of each node with $\omega_i=\omega=3$. We vary the interlayer coupling strength $k_R$ from $0.01$ to $-0.01$ with fixed space $-0.0001$ and fixed intralayer coupling strength $k_A=0.1$. For each of these $200$ $k_R$s, we choose the initial condition of each oscillator randomly within the interval $[-1,1] \times [-1,1]$. For $k_R>0$, the system remains in interlayer phase synchronization (i.e., $F_{Replica}=2$) beyond a critical value of $k_R$. However, the system attains interlayer antisynchronization $(F_{Replica}=0)$ for a suitable negative interlayer coupling strength. Each of these subfigures is drawn with different adjacency matrices $\mathscr{\tilde{A}}^{[\alpha]}$ using the multiplex network in Fig.\ \eqref{fig1}. Subfigure (a) represents the results for hub-attracting intralayer matrix $(\beta=1)$, whereas the subfigure (c) depicts the results for the hub-repelling intralayer matrix $(\beta=-1)$. The middle panel (subfigure (b)) contemplates the results for the unweighted intralayer matrix $(\beta=0)$. Irrespective of the chosen value of $\beta$, the system settles down to an interlayer antisynchronized state for negative interlayer coupling strength (See blue square markers). {\color{black}In spite of choosing negative $k_R$, each layer maintains intralayer synchronization as $F_{Layer1}$ (red plus (+) markers) $=F_{Layer2}$ (magenta circle markers) $=2$ throughout the subfigures.}
	}\label{fig2}
\end{figure*}

Here, $c_1$ is the integration constant. Also, the linear stability analysis of \eqref{19} suggests there are two stationary points, viz.\ (i) $r_1=0$, and (ii) $r_1=1$. The stationary point $r_1=0$ is unstable. In fact, we are not interested in $r_1=0$, as $r_1=0$ corresponds to the amplitude death state, which contradicts the emergence of the interlayer antisynchronization state. The other stationary point $r_1=1$ is stable. Thus, the system \eqref{14} experiencing the intralayer synchronization and the interlayer antisynchronization possesses the unit amplitude $(r_1=1)$ irrespective of the choice of the coupling coefficients $k_A$ and $k_R$.

\subsubsection{Numerical illustration and Demultiplexing effect}

\begin{figure*}[ht]
	\centerline{\includegraphics[scale=0.5]{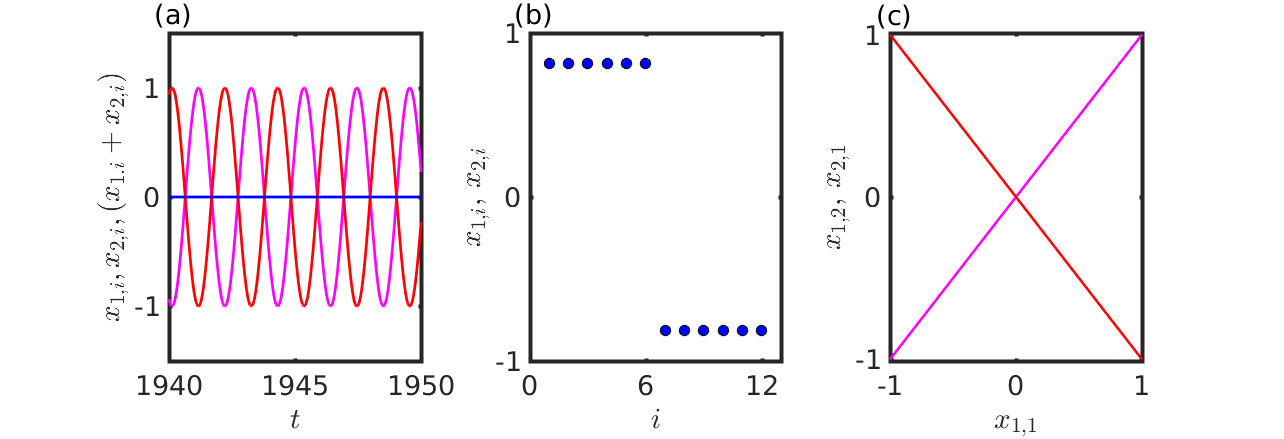}}
	\caption{{\bf Intralayer synchronization and interlayer antisynchronization for hub-attracting intralayer matrix $(\beta=1)$}: All the trajectories of the first layer collapse to a single trajectory (red line), and similarly, the trajectories of the second layer oscillate within $[-1,1]$ maintaining the same path (magenta line) in subfigure (a). This attests to the occurrence of intralayer synchronization. The sum $(x_{1,i}+x_{2,i})$ converges to a fixed value zero (blue line) after the transient. This validates the emergence of interlayer antisynchronization. Subfigure (b) contemplates the appearance of two clusters. The SL oscillators of the first layer lie within a synchronized group, and the oscillators of the second layer stay in another cluster. Due to the presence of repulsive interlayer coupling strength $k_R=-0.1$, these two clusters maintain a constant phase difference of $\pi$. All the subfigures are drawn using random initial conditions from $[-1,1] \times [-1,1]$. We choose the oscillators on top of the node-$1$, $2$, and $7$ respectively, from the multiplex network given in Fig.\ \eqref{fig1}. We plot the phase portrait of these oscillators after the transient. Clearly, we have $x_{1,1}=x_{1,2}=-x_{2,1}$. More importantly, subfigures (a) and (c) confirm our analytical calculation revealing each SL oscillator evolves with a unit radius after reaching the interlayer antisynchronization manifold and the intralayer synchronization manifold. For each subfigure, we choose $k_A=|k_R|=0.1$. 
	}\label{fig3}
\end{figure*}

\par To validate our analytical findings, we consider the multiplex network given in Fig.\ \eqref{fig1}. This multiplex network contains two layers, where each layer consists of two different connected intralayer networks. The first layer contains $6$ nodes and $7$ links, whereas the second layer is made of $6$ nodes and $8$ edges. On top of each of these vertices, we place identical SL oscillators \eqref{13} with the same intrinsic frequency $\omega=3.0$. To verify our findings, here we propose two different measures, viz.\ 

\begin{enumerate}
	\item The first one 
	
	\begin{equation}\label{21}
	\begin{split}
	F_{Replica}={\Big\langle\frac{1}{N}\sum_{i=1}^{N} [1+\cos{({\theta}_{1,i}-{\theta}_{2,i})}]\Big\rangle}
	\end{split}
	\end{equation}
	
	is to measure the interlayer antisynchronization. ${\langle \cdot \rangle}$ represents here the time average, and for numerical simulation, we choose $0.5 \times 10^5$ steps to average this measure after the initial transients of $1.5 \times 10^5$ steps. The scaling factor $\frac{1}{N}$ accounts for the $N$ number of interlayer links. We are basically interested with only two values of $F_{Replica}$, viz.\ $F_{Replica}=2$, which indicates the interlayer phase synchronization, and $F_{Replica}=0$ representing the interlayer antisynchronization. However, this measure deals with only the phase of each oscillator; thus, to ensure the intralayer synchronization and interlayer antisynchronization, we need to see the temporal evolution of the state vectors too. 
	
	\item 
	
	To measure intralayer phase synchronization, we define
	
	\begin{equation}\label{22}
	\begin{split}
	F_{Layer1}={\Big\langle\frac{1}{L_{1}}\sum_{i<j}\mathscr{A}_{ij}^{[1]}[1+\cos{({\theta}_{1,i}-{\theta}_{1,j})}] \Big\rangle},\\
	F_{Layer2}={\Big\langle\frac{1}{L_{2}}\sum_{i<j}\mathscr{A}_{ij}^{[2]}[1+\cos{({\theta}_{2,i}-{\theta}_{2,j})}] \Big\rangle}.
	\end{split}
	\end{equation}
	
	Here, $L_1$ and $L_2$ are the numbers of edges of both connected layers, respectively. If these two measures attain their respective maximum values of $2$, the system achieves intralayer phase synchronization. Besides, if they both acquire their respective minimum values $0$, the system reaches intralayer antiphase synchronization. 
	
\end{enumerate}

Using the multiplex network in Fig.\ \eqref{fig1}, we have construct the adjacency matrix $ \mathscr{A}$ (See \eqref{4}). The degree of each node is given by $d_1=d_5=d_7=d_9=d_{11}=d_{12}=4$ and $d_2=d_3=d_4=d_6=d_8=d_{10}=3$. Using these degrees $d_i$ and intralayer graphs, we construct the weighted directed networks with adjacency matrices $\mathscr{\tilde{A}}^{[\alpha]}$, $\alpha=1,2$. In Fig.\ \eqref{fig2}, we plot the variation of $F_{Layer1}$, $F_{Layer2}$, and $F_{Replica}$ by numerically integrating Eqs.\ \eqref{2} with intralayer coupling strength $k_A=0.1$. For all the numerical simulations with identical SL oscillators, we choose initial conditions randomly for each oscillator within the interval $[-1,1] \times [-1,1]$. An exciting observation of Fig.\ \eqref{fig2} is that the system does not exhibit interlayer antisynchronization for any positive interlayer coupling coefficient $k_R$. Once the interlayer coupling strength $k_R$ becomes negative and attains a sufficient value, $F_{Replica}$ diminishes to zero and continues to be at zero, suggesting the occurrence of interlayer antisynchronization. To compare the results, we vary $k_R$ within the interval $\big[-\frac{k_A}{10},\frac{k_A}{10}\big]$ in each subfigures, where $k_A=0.1$. We vary $k_R$ from $0.01$ to $-0.01$ with small space $-0.0001$, and for each step, we select the initial conditions randomly from $[-1,1] \times [-1,1]$. The subfigures (a-c) are plotted for the hub-attracting intralayer matrix $(\beta=1)$, the unweighted intralayer matrix $(\beta=0)$, and the hub-repelling intralayer matrix $(\beta=-1)$, respectively. Depending on the initial conditions in the small neighborhood of $k_R=0$, $F_{Replica}$ attains multiple values. However, the measures $F_{Layer1}$ and $F_{Layer2}$ reach their maximum values of $2$ for all chosen values of the interlayer coupling strength $k_R$, even when $k_R$ is negative. This suggests our chosen intralayer coupling strength $k_A$ for this simulation is sufficient to maintain the coherent behavior among the identical SL oscillators within the layers, and the interlayer coupling strength $k_R$, even when it is negative, can not destroy the intralayer coherence. Nevertheless, for all these three matrices, $F_{Replica}$ becomes zero beyond a critical value of $k_R<0$. The required interlayer coupling strength with fixed $k_A=0.1$ for the multiplex network given in Fig.\ \eqref{fig1} is as follows:
(i) $k_R \approx -0.0037$ for hub-attracting intralayer matrix $(\beta=1)$,
(ii) $k_R \approx -0.0033$ for unweighted intralayer matrix $(\beta=0)$, and
(iii) $k_R \approx -0.0032$ for hub-repelling  intralayer matrix $(\beta=-1)$. All these critical values are obtained after averaging over $100$ independent numerical simulations. We have the same underlying network structure in all these realizations but possess different random initial conditions. Thus, the critical interlayer coupling strength varies with each realization. Note that we do not want to emphasize the role of enhancement of interlayer antisynchronization here by changing the values of $\beta$. This topic is a subject of rigorous investigation and beyond the scope of the present work. The impact of $\beta$ on the enhancement of interlayer antisynchronization and determine the critical value of $k_R$ for different multiplex networks may be investigated in the near future. Nevertheless, we later perform the global stability analysis of the interlayer antisynchronization for a few special intralayer networks to elucidate the effect of initial conditions. i.e., {\color{black}we will determine an approximate 
	 value of interlayer coupling strength $k_R<0$ for which the system evolves interlayer antisynchronously irrespective of the choice of initial conditions, except for a set of measure zero.}

\par Thus, our selected intralayer coupling function $G$, interlayer coupling function $H$, intralayer coupling strength $k_A>0$, and intralayer coupling strength $k_R<0$ work immensely well for the emergence of intralayer synchronization and interlayer antisynchronization states. However, as mentioned earlier, the proposed measures do not incorporate the amplitude of the oscillators. Hence, we plot the dynamics of each SL oscillator in Fig.\ \eqref{fig3}. To avoid monotonicity, we show the results in Fig.\ \eqref{fig3} with only hub-attracting intralayer matrix $(\beta=1)$. Although we plot all twelve oscillators' temporal evolution in subfigure (a) of Fig.\ \eqref{fig3}, however, we can only see two trajectories in this subfigure. This is due to the simultaneous appearance of intralayer synchronization in both layers. All trajectories of the same layer collapse into a single one. To generate this figure, we choose the same multiplex network given in Fig.\ \eqref{fig1}. We set the intralayer coupling strength $k_A=0.1$ and the intralayer coupling strength $k_R=-0.1$, so that $|k_A|=|k_R|$. We again choose initial conditions randomly for each SL oscillator within the interval $[-1,1] \times [-1,1]$. Interestingly, the two intralayer synchronized trajectories maintain a constant $\pi$ phase difference as revealed through Fig.\ \eqref{fig3} (a). Apart from that, we also plot the sum $(x_{1,i}+x_{2,i})$ (See blue line) that indicates the sum of the dynamics of the oscillators situated on top of the replica nodes. This temporal evolution of $(x_{1,i}+x_{2,i})$ remains at zero after the initial transient as depicted in the subfigure (a) of Fig.\ \eqref{fig3}. This $(x_{1,i}+x_{2,i})=0$ suggests the emergence of interlayer antisynchronization. We also plot the positions of $x_{1,i}$ and $x_{2,i}$ of the multiplex at a particular time after the initial transient in subfigure (b). This snapshot indicates the occurrence of two synchronized clusters with $\pi$ phase difference. Evidently, one of these synchronized clusters represents the state of the oscillators of one layer, and the other cluster reflects the dynamics of another layer. Subfigure (c) reveals any oscillator of the first layer (here, without any loss of generality, we choose $x_{1,1}$, i.e., the first oscillator) maintains a synchronized rhythm with any oscillator of the same layer (here, we choose $x_{1,2}$, i.e., the second oscillator for visualization) and preserves the interlayer antisynchronization with the oscillators on top of the replica node of the other layer (See $x_{1,1}=-x_{2,1}$ in Fig.\ \eqref{fig3} (c)). Further, these two subfigures (a) and (c) of Fig.\ \eqref{fig3} ensure that the identical SL oscillators sustain a unit radius after achieving the intralayer synchronization and interlayer antisynchronization. This validates our analytical findings too.

\begin{widetext}
	\hspace{0.5cm}
	\begin{table*}[h!]
		\begin{center}
			\caption{Demultiplexing of the network}
			\label{Table1}
			\begin{tabular}{||c | c | c | c||}  
				\hline
				Removal of interlayer links & $F_{Layer1}$ & $F_{Layer2}$ & $F_{Replica}$ \\ [0.5ex] 
				\hline\hline
				$1-7$ & 2 & 2 & 0 \\ 
				\hline
				$2-8$ & 2 & 2 & 0 \\
				\hline
				$3-9$ & 2 & 2 & 0 \\
				\hline
				$4-10$ & 2 & 2 & 0 \\
				\hline
				$5-11$ & 2 & 2 & 0 \\
				\hline
				$6-12$ & 2 & 2 & Initial conditions dependent value \\ [1ex] 
				\hline
			\end{tabular}\label{Table}
		\end{center} 
	\end{table*}
\end{widetext}

\begin{figure*}[ht]
	\centerline{\includegraphics[scale=0.5]{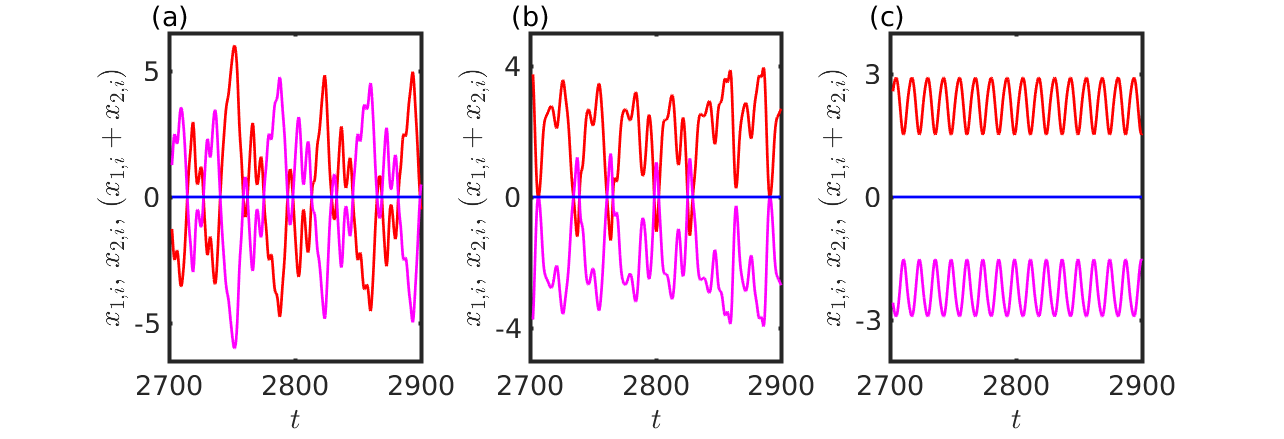}}
	\caption{{\bf The intralayer synchronization and intralayer antisynchronization with identical Thomas' cyclically symmetric attractor}: The figures are drawn for (a) $b=0.10$, (b) $b=0.20$, and (c) $b=0.30$. We consider the same multiplex network with $12$ vertices and $21$ edges, shown in Fig.\ \eqref{fig1}. Initial conditions are chosen randomly from $[-4,4] \times [-4,4] \times [-4,4]$. The interlayer coupling strength $k_R$ is set at $-0.3$, and the intralayer coupling strength $k_A$ is kept fixed at $1.0$. We choose $\beta=-1$; thus, we have the hub-repelling intralayer matrices. In all these subfigures, we find $(x_{1,i}+x_{2,i})$ converges to a fixed value zero after the initial transient. Hence, the emergence of interlayer antisynchronization is confirmed. Moreover, all the trajectories of the same layer collapse into a single trajectory (shown in the red (magenta) line for the first (second) layer). 
	}\label{fig4}
\end{figure*}

\par Now, we want to understand whether all these $N=6$ interlayer links are necessary or not to achieve interlayer antisynchronization in the multiplex, chosen in Fig.\ \eqref{fig1}. Instead of demultiplexing the multiplex randomly, we prefer a systematic way to demultiplex the network. First, we remove the connections between the first oscillator of both layers, i.e., we disconnect the interlayer link $1-7$ of the multiplex shown in Fig.\ \eqref{fig1}. Now, in the absence of this link $1-7$, the network does not remain as a multiplex. However, it remains a multilayer network. Now, we integrate the system \eqref{2} by placing identical SL oscillators \eqref{13} on top of each node with $\omega=3$. We choose the interlayer coupling strength $k_R=-0.1$ and set the intralayer coupling strength as $k_A=|k_R|$. We again choose the initial conditions randomly within $[-1,1] \times [-1,1]$. The chosen coupling strengths still allow the system to maintain the interlayer antisynchronization along with intralayer synchronization. Keeping the same coupling strengths and random initial conditions from $[-1,1] \times [-1,1]$, we remove the link $2-8$ between the second oscillators of both layers. Interestingly, even this link removal does not destroy both the interlayer antisynchronization and intralayer synchronization. In fact, in this way, we gradually disconnect the interlayer links one by one. We find the system evolves in the interlayer antisynchronization and intralayer synchronization; still, there exists at least one interlayer link between the two layers. Unless we detach the last interlayer link $6-12$, i.e., the connection between the sixth oscillators of both layers, the system settles in the interlayer antisynchronization state. Thus, only one interlayer link is sufficient to entertain the interlayer antisynchronization once the oscillators settle themselves into the intralayer synchronization manifold.

{\color{black}
\par Here is a feasible explanation behind this occurrence of interlayer antisynchronization with only one interlayer link. Once the oscillators attain intralayer synchronization, this coherence will not be destroyed with negative interlayer coupling strength, as shown in Figs.\ \eqref{fig2} and \eqref{fig3}. Thus, each of these two layers can be represented by two state vectors ${\bf x}_{1}(t)$ and ${\bf x}_{2}(t)$ (say), respectively. Now, since that one single interlayer link connects these two layers with repulsive interlayer coupling strength, thus these two state vectors ${\bf x}_{1}(t)$ and ${\bf x}_{2}(t)$  try to maximize their phase difference. Hence, we have $|{\bf x}_{1}(t)|=|{\bf x}_{2}(t)|$ and their phase difference is exactly $\pi$. In other words, we have ${\bf x}_{1}(t)+{\bf x}_{2}(t)={\bf 0}$. In the table \eqref{Table}, we represent how the gradual removal of interlayer links results in the values of $F_{Layer1}$, $F_{Layer2}$, and $F_{Replica}$. We find $F_{Replica}=0$ until the single interlayer link $6-12$ remains. After removing all interlayer links, $F_{Replica}$ will give an initial condition-dependent value. Deleting all interlayer links still entertains the intralayer synchronization as the intralayer coupling strength $k_A=0.1$ provides sufficient coherence among the oscillators within the same layer. Thus, we have $F_{Layer1}=F_{Layer2}=2$ even without all interlayer connections.
}

\subsection{Thomas' cyclically symmetric attractor}

\par It is already established in Sec.\ \ref{Analytical} that the vector field $f$ should be an odd function in order to realize one of the necessary conditions of interlayer antisynchronization along with intralayer synchronization. We already represent the results with the help of SL oscillators in the earlier subsection. To further validate our claim, we choose a different system with a self-excited attractor \cite{dudkowski2016hidden,nag2020hidden}, viz. Thomas' cyclically symmetric attractor \cite{thomas1999deterministic,sprott2007labyrinth,rowlands2008simple}, where the state dynamics of the $i$-th oscillator is represented by

\begin{equation} \label{23}
f({\bf x}_i)=\left(
\begin{array}{c}
\sin{(y_i)}-bx_i\\
\sin{(z_i})-by_i\\
\sin{(x_i)}-bz_i\\
\end{array}
\right), \\
\end{equation}

where $b$ is a constant. For $b>1$, the origin is the single stable equilibrium. The system undergoes a pitchfork bifurcation at $b=1$. As the parameter $b$ is further decreased, the system undergoes a Hopf bifurcation around $b \approx 0.32899$, creating stable limit cycles. Through a period-doubling cascade, the system becomes chaotic at $b \approx 0.208186$. 


\par We integrate Eqs.\ \eqref{5} with $\beta=-1$ by placing identical Thomas' cyclically symmetric attractor on top of each node. We simulate the system for $3 \times 10^5$ steps and discard the initial $2.7 \times 10^5$ steps treating them as transient. We choose the same coupling functions $G({\bf x}_i,{\bf x}_j)=[{x}_j - {x}_i, {y}_j - {y}_i, {z}_j - {z}_i]^T$ and $H({\bf x}_i,{\bf x}_j)=[{x}_j + {x}_i, {y}_j + {y}_i, {z}_j + {z}_i]^T$ for the numerical simulation. All the subfigures in Fig.\ \eqref{fig4} are drawn with fixed $k_A=1.0$ and $k_R=-0.3$. We choose three distinct values of the system parameter $b$. All these subfigures suggest all the six oscillators of the same layer coincide in a single trajectory, indicating the intralayer synchronization. However, they exhibit replica-wise antiphase synchronization. We find $(x_{1,i}+x_{2,i})$ (blue line) converges to exactly zero after the initial transient in Fig.\ \eqref{fig4}. Hence, we again confirm the emergence of the interlayer antisynchronization with Thomas' cyclically symmetric attractor and validate our analytical calculations for such a state's existence.

\section{Local stability analysis of interlayer antisynchronization state}

\par We already derive a necessary condition

\begin{equation}\label{24}
\begin{split}
\sum_{j=1}^{N}\mathscr{\tilde{A}}_{ij}^{[1]} G[{\bf x}_{1,j},{\bf x}_{1,i}]=-\sum_{j=1}^{N}\mathscr{\tilde{A}}_{ij}^{[2]} G[-{\bf x}_{1,j},-{\bf x}_{1,i}]
\end{split}
\end{equation}

for the emergence of interlayer antisynchronization. Still, now, we ignore this condition as the function $G$ is chosen as the diffusive function and it will vanish identically after the occurrence of the intralayer synchronization. Therefore, the condition \eqref{24} mentioned above is trivially satisfied. Since, $G$ is chosen as $G({\bf x}_i,{\bf x}_j)=[{x}_j - {x}_i, {y}_j - {y}_i, {z}_j - {z}_i]^T=G[{\bf x}_i-{\bf x}_j]$ (say). Thus from Eq.\ \eqref{24}, we have 

\begin{equation}\label{25}
\begin{split}
\sum_{j=1}^{N}\mathscr{\tilde{A}}_{ij}^{[1]} G[{\bf x}_{1,j}-{\bf x}_{1,i}]=-\sum_{j=1}^{N}\mathscr{\tilde{A}}_{ij}^{[2]} G[-{\bf x}_{1,j}+{\bf x}_{1,i}],\\
\Rightarrow \sum_{j=1}^{N}\mathscr{\tilde{A}}_{ij}^{[1]} G[{\bf x}_{1,j}-{\bf x}_{1,i}]=\sum_{j=1}^{N}\mathscr{\tilde{A}}_{ij}^{[2]} G[{\bf x}_{1,j}-{\bf x}_{1,i}].
\end{split}
\end{equation}

The interlayer antisynchronization and the intralayer synchronization states are two completely independent emerging phenomena of a multiplex network. Thus, the system does not need to evolve into intralayer synchrony during the appearance of interlayer antisynchronization. Therefore, $G$ may not vanish during the sole occurrence of interlayer antisynchronization. Hence for $G \neq 0$, we have from Eq.\ \eqref{25} the following simplistic choice 

\begin{equation}\label{26}
\begin{split}
\mathscr{\tilde{A}}^{[1]}=\mathscr{\tilde{A}}^{[2]}=\mathscr{\tilde{B}} \text{(say)}.
\end{split}
\end{equation}

Thus, both the connected layers contain the same intralayer networks. Hence, Eqs.\ \eqref{2} transform to the following set of equations

\begin{equation} \label{eq_27}
\begin{array}{lll}
\dot{\bf x}_{1,i}=f({\bf x}_{1,i})+k_A\sum\limits_{j=1}^{N}\mathscr{\tilde{B}}_{ij}G[{\bf x}_{1,j}-{\bf x}_{1,i}]\\+k_R H[{\bf x}_{2,i}+{\bf x}_{1,i}],\\[5pt]
\dot{\bf x}_{2,i}=f({\bf x}_{2,i})+k_A\sum\limits_{j=1}^{N}\mathscr{\tilde{B}}_{ij}G[{\bf x}_{2,j}-{\bf x}_{2,i}]\\+k_R H[{\bf x}_{2,i}+{\bf x}_{1,i}],
\end{array}
\end{equation}

During the occurrence of interlayer antisynchronization state, the synchronous solution satisfies 

\begin{equation} \label{eq_28}
\begin{array}{lll}
\dot{\bf x}_{1,i}=f({\bf x}_{1,i})+k_A\sum\limits_{j=1}^{N}\mathscr{\tilde{B}}_{ij}G[{\bf x}_{1,j}-{\bf x}_{1,i}],\\[5pt]
\dot{\bf x}_{2,i}=-\dot{\bf x}_{1,i},
\end{array}
\end{equation}

Let, $\delta{\bf \chi}_i(t)$ be a tiny amount of feasible perturbation on the $i$-th node of the second layer from its interlayer antisynchronization state. Then, we have

\begin{equation} \label{eq_29}
\begin{array}{lll}
{\bf x}_{2,i}(t)=-{\bf x}_{1,i}(t)+\delta{\bf \chi}_i(t).
\end{array}
\end{equation}

Thus, the error dynamics transverse to the interlayer antisynchronization manifold $\Xi=\big\{ \big({\bf x}_{1,1}(t),{\bf x}_{1,2}(t),\dots,{\bf x}_{1,N}(t)\big)\subseteq \mathbb{R}^{mN}~:~{\bf x}_{1,i}(t)+{\bf x}_{2,i}(t)={\bf 0}$ for all $i=1,2,\dots,N~\mbox{and}~t\in\mathbb{R}^+ \big\}$ is given by the following equations

\begin{equation} \label{eq_30}
\begin{array}{lll}
\delta\dot{\bf \chi}_i&=\dot{\bf x}_{1,i}+\dot{\bf x}_{2,i}\\[5pt]
&=f({\bf x}_{1,i})+f(-{\bf x}_{1,i}+\delta{\bf \chi}_i)\\&~~~+k_A \sum\limits_{j=1}^{N}\mathscr{\tilde{B}}_{ij}G[\delta{\bf \chi}_j-\delta{\bf \chi}_i]+2k_R H\delta{\bf \chi}_i,\\[5pt]
&=Jf({\bf x}_{1,i})\delta{\bf \chi}_i-k_A\sum\limits_{j=1}^{N}\mathscr{\tilde{L}}_{ij}G\delta{\bf \chi}_j+2k_R H\delta{\bf \chi}_i,
\end{array}
\end{equation}

for all $i=1,2,\dots,N$. Here, $Jf({\bf x}_{1,i})=\dfrac{\partial f({\bf x})}{\partial{\bf x}}\Big|_{{\bf x}={\bf x}_{1,i}}$, where ${\bf x}_{1,i}$  satisfies Eq.\ \eqref{eq_28}. Also, $\mathscr{\tilde{L}}$ be the zero-row sum intralayer Laplacian matrix \cite{anwar2021enhancing}, defined as $\mathscr{L}_{ij}=-\mathscr{B}_{ij}$ for $i\ne j$ and $\mathscr{L}_{ii}=\sum_{j=1}^{N}\mathscr{B}_{ij}$, $i=1,2,3,\cdots,N$. Due to the linear independence of these error components, all the state variables of Eq.\ \eqref{eq_30} evolve transverse to the interlayer antisynchronization manifold. Therefore, the Lyapunov exponents of the Eq. \eqref{eq_30} are all transverse to $\Xi$.

\par Now we place Thomas cyclically symmetric attractor on top of each node of the multiplex, and thus, using Eq.\ \eqref{eq_30}, we derive the following transverse error equation

\begin{equation}\label{eq_31}
\begin{array}{lll}
\delta\dot{x}_i=cos(y_i)\delta y_i-b\delta x_i-k_A\sum\limits_{j=1}^{N}\mathscr{\tilde{L}}_{ij}\delta x_j+2k_R\delta x_i,\\[10pt]
\delta\dot{y}_i=cos(z_i)\delta z_i-b\delta y_i-k_A\sum\limits_{j=1}^{N}\mathscr{\tilde{L}}_{ij}\delta y_j+2k_R\delta y_i,\\[10pt]
\delta\dot{z}_i=cos(x_i)\delta x_i-b\delta z_i-k_A\sum\limits_{j=1}^{N}\mathscr{\tilde{L}}_{ij}\delta z_j+2k_R\delta z_i.
\end{array}
\end{equation}

\par For each $i=1,2,\cdots,N$, the state variable $(x_i,y_i,z_i)$, being lying on the interlayer antisynchronization manifold, satisfies the following equations

\begin{equation}\label{eq_32}
\begin{array}{lll}
\dot{x}_i=cos(y_i)-bx_i-k_A \sum\limits_{j=1}^{N}\mathscr{\tilde{L}}_{ij}x_j,\\[10pt]
\dot{y}_i=cos(z_i)-by_i-k_A\sum\limits_{j=1}^{N}\mathscr{\tilde{L}}_{ij}y_j,\\[10pt]
\dot{z}_i=cos(x_i)-bz_i-k_A\sum\limits_{j=1}^{N}\mathscr{\tilde{L}}_{ij}z_j.
\end{array}
\end{equation}

\begin{figure}[ht]
	\centerline{\includegraphics[scale=0.35]{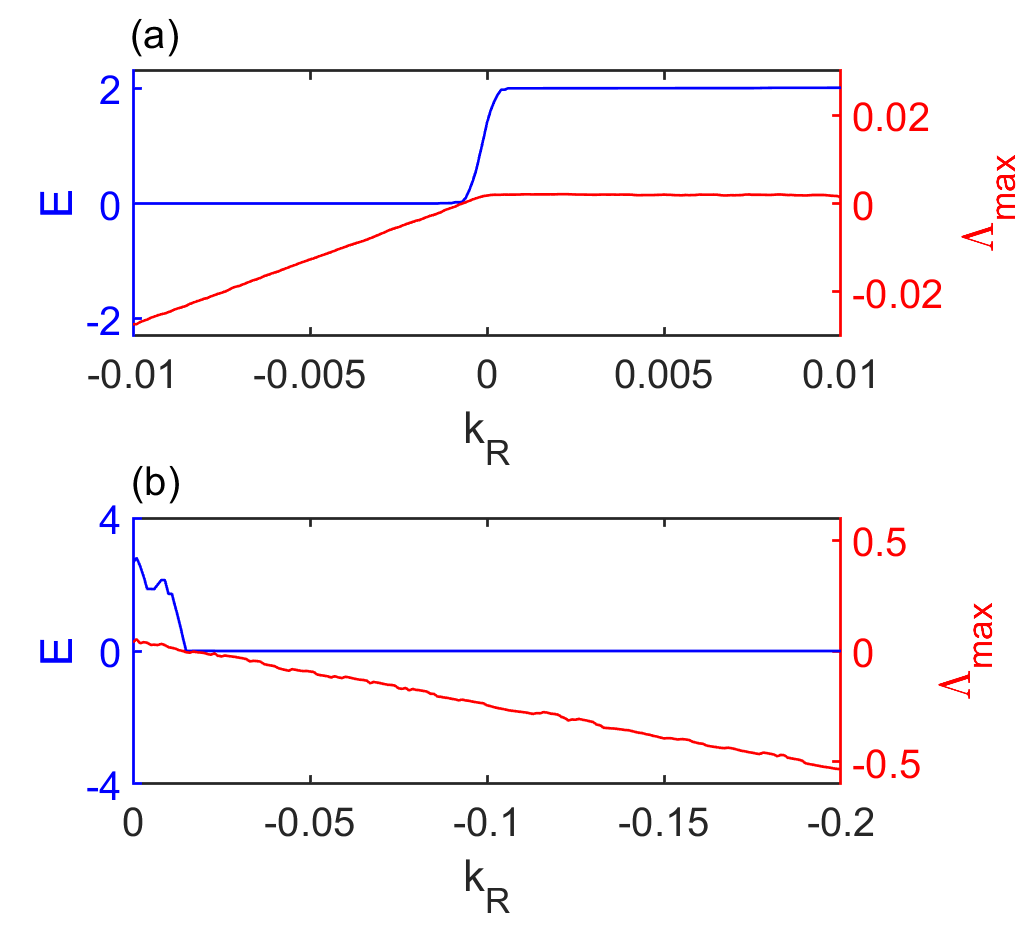}}
	\caption{{\bf The interlayer antisynchronization error $E$ (blue) and the maximum transverse Lyapunov exponent $\Lambda_{max}$ (red) as a function of the interlayer coupling strength $k_R$}: Both layers contain the same ring intralayer network with four vertices. In subfigure (a) for the SL oscillator, we set the system's parameter at $k_A=0.1$ and $\omega=3.0$. Whereas, in subfigure (b) for the Thomas cyclically symmetric attractor, we set $k_A=1.0$ and $b=0.2$. $E$ reduces to zero suggesting the occurrence of interlayer antisynchronization. Simultaneously, $\Lambda_{max}$ crosses zero and becomes negative, revealing the appearance of interlayer antisynchronization. This indicates that our local stability condition agrees quite well with our numerical simulation.
	}\label{fig5}
\end{figure}

\par Since we are interested in investigating the local stability of the interlayer
antisynchronization state for the duplex network of Thomas cyclically symmetric attractor, we calculate $3N$ Lyapunov exponents by solving the linearized equation \eqref{eq_31} along with the equation of motion \eqref{eq_32} of the interlayer antisynchronization state. Out of these $3N$ Lyapunov exponents, the maximum Lyapunov exponent $\Lambda_{max}$ will provide us the transition point from desynchronization to interlayer antisynchronization state. By keeping fixed the intralayer coupling strength $k_A$, we plot $\Lambda_{max}$ as a function of the interlayer coupling strength $k_R$. $\Lambda_{max}<0$ will provide the necessary condition for the local stability of the interlayer antisynchronization state.

\par Similarly, if we place identical SL oscillators on top of each node of the multiplex instead of Thomas cyclically symmetric attractor, the error components transverse to the interlayer antisynchronization manifold satisfy the following evolution equation

\begin{equation}\label{eq_50}
\begin{array}{lll}
\delta\dot{x}_i=[1-3{x}^2_i-y^2_i]\delta x_i-[\omega+2x_iy_i]\delta y_i\\-k_A\sum\limits_{j=1}^{N}\mathscr{\tilde{L}}_{ij}\delta x_j+2k_R\delta x_i,\\[10pt]
\delta\dot{y}_i=(\omega-2x_iy_i)\delta x_i+[1-x^2_i-3y^2_i]\delta y_i\\-k_A\sum\limits_{j=1}^{N}\mathscr{\tilde{L}}_{ij}\delta y_j+2k_R\delta y_i.
\end{array}
\end{equation}

For each $i=1,2,\cdots,N$, the state variable $(x_i,y_i,z_i)$ of the interlayer antisynchronization state satisfies the following equations

\begin{equation}\label{eq_51}
\begin{array}{lll}
\dot{x}_i=[1-(x^2_i+y^2_i)]x_i-\omega y_i-k_A \sum\limits_{j=1}^{N}\mathscr{\tilde{L}}_{ij}x_j,\\[10pt]
\dot{y}_i=[1-(x^2_i+y^2_i)]y_i+\omega x_i-k_A\sum\limits_{j=1}^{N}\mathscr{\tilde{L}}_{ij}y_j.
\end{array}
\end{equation}

For the computation of the maximum Lyapunov exponent $\Lambda_{max}$ for the duplex of SL oscillators, we need to solve the linearized Eq.\ \eqref{eq_50} along with Eq.\ \eqref{eq_51}  of the interlayer antisynchronization state, yielding the $2N$ Lyapunov exponents.

\par Figure \eqref{fig5} indicates our derived local stability condition works quite well. We choose the intralayer coupling strength $k_A=0.1$ and the system parameter $\omega=3.0$ for the SL oscillator in Fig.\ \eqref{fig5} (a). Whereas we keep fixed $k_A=1.0$,
and the system parameter $b=0.2$ for the Thomas cyclically symmetric attractor in Fig.\ \eqref{fig5} (b). We choose a ring network of four nodes in each layer for this simulation. As the constructed multiplex is a regular graph where the degree of each vertex of the multiplex is three, thus we have $\mathscr{\tilde{A}}=\mathscr{A}$ from Eq.\ \eqref{3}. We plot the interlayer antisynchronization error $E$ (See Eq.\ \eqref{7}) in Fig.\ \eqref{fig5} for both coupled systems. Clearly, $E$ (blue) diminishes to zero gradually with the decrement of $k_R$. This $E=0$ attests to the emergence of the interlayer antisynchronization state ${\bf x}_{1,i}(t)+{\bf x}_{2,i}(t)={\bf 0}$, $\forall \hspace{0.2cm} i=1,2,3,4$. Similarly, the red lines in Fig.\ \eqref{fig5} contemplate the variation of the maximum Lyapunov exponent $\Lambda_{max}$ as a function of $k_R$. As evident from Fig.\ \eqref{fig5}, $\Lambda_{max}$ becomes negative where $E$ becomes zero in Fig.\ \eqref{fig5}.

{\color{black}
\section{Sufficient condition of global stability analysis for interlayer antisynchronization state}}

\par To derive the global stability condition for the interlayer antisynchronization state, we need to assume a few conditions which need to be satisfied for our calculations. If the following conditions are satisfied, our derived interlayer coupling strength $k_R$ leads to the interlayer antisynchronization in the duplex network irrespective of the chosen initial conditions except for a set of measure zero.

\begin{enumerate}
	\item The first condition is the individual vector field $f$ must be Lipschitz continuous, i.e., there exists a positive real constant $M$ such that 
	
	\begin{equation}\label{eq_33}
	\begin{array}{lll}
	\dfrac{\norm{f({\bf x})-f({\bf y})}}{\norm{{\bf x}-{\bf y}}} \le M, \hspace{0.2cm}{\bf x} \neq {\bf y}.
	\end{array} 
	\end{equation}
	
	If the relation mentioned above holds, i.e., if there is an upper bound of the rate of change of the isolate oscillators' dynamics in the phase space, then capitalizing on the Cauchy-Schwarz inequality, we have

	\begin{equation}\label{eq_34}
	\begin{array}{lll}
	[{\bf x}-{\bf y}]^{T}[f({\bf x})-f({\bf y})] \le\norm{{\bf x}-{\bf y}}\norm{f({\bf x})-f({\bf y})} \\ \hspace{3.5cm}\le M({\bf x}-{\bf y})^{T}({\bf x}-{\bf y})\\ \hspace{5cm}\forall \hspace{0.2cm} {\bf x},{\bf y}\in\mathbb{R}^m
	\end{array} 
	\end{equation}
	
	\item The intralayer Laplacian matrix $\mathscr{\tilde{L}}$ must be a symmetric matrix. However, as per our construction, $\mathscr{\tilde{L}}$ is a symmetric matrix if and only if the intralayer network is a regular graph {\color{black}(i.e., $d_i=d_j$ for all $i$ and $j$)}, or we choose $\beta=0$, i.e., we consider only the unweighted case. Thus, we restrict our global stability analysis to two types of intralayer networks, viz.\ $\beta=0$ or the regular intralayer networks where each node has the same degree. Hence under these two specific choices, the intralayer Laplacian matrix $\mathscr{\tilde{L}}$ is a symmetric positive semi-definite matrix. Thus, one of its eigenvalues is zero, and all the other eigenvalues are positive.
	
	\item The interlayer coupling matrix $H$ must be a symmetric positive definite. Thus all of its eigenvalues are strictly positive.
	
	\item The intralayer coupling matrix $G$ is a symmetric positive semi-definite matrix. Therefore all the eigenvalues of $G$ are non-negative.

\end{enumerate}

\par Let us define the interlayer antisynchronization error for each replica as follows

\begin{equation}\label{eq_35}
\begin{array}{lll}
{\bf e}_i={\bf x}_{1,i}+{\bf x}_{2,i}$ for $i=1,2,\cdots,N.
\end{array} 
\end{equation}

Hence, we have

\begin{equation} \label{eq_36}
\begin{array}{lll}
\dot{\bf e}_i=\dot{\bf x}_{1,i}+\dot{\bf x}_{2,i}\\[5pt]
=f({\bf x}_{1,i})+f(-{\bf x}_{1,i}+{\bf e}_i)+k_A\sum\limits_{j=1}^{N}\mathscr{\tilde{B}}_{ij}G[{\bf e}_j-{\bf e}_i]\\+2k_R H{\bf e}_i,\\[5pt]
=f({\bf x}_{1,i})-f({\bf x}_{1,i}-{\bf e}_i)-k_A\sum\limits_{j=1}^{N}\mathscr{\tilde{L}}_{ij}G{\bf e}_j+2k_R H{\bf e}_i.
\end{array}
\end{equation}

Let ${\bf e}$ be the stack of the error terms ${\bf e}_1,{\bf e}_2,\dots,{\bf e}_N$ in the vectorial form. Then, we can rewrite the rate of change of this error system in the following form,

\begin{equation} \label{eq_37}
\begin{array}{lll}
\dot{\bf e}=\bigoplus\limits_{i=1}^{N}\big[f({\bf x}_{1,i})-f({\bf x}_{1,i}-{\bf e}_i)\big]-k_A\mathscr{\tilde{L}}\otimes G{\bf e}\\+2k_R I_N\otimes H{\bf e},
\end{array}
\end{equation}

where ${\bf x}_{1,i},{\bf x}_{2,i}$ satisfy Eq.\ \eqref{eq_27}. Here, $\bigoplus$ and $\otimes$ represent the matrix direct sum and Kronecker product, respectively. 

\par Let us define a Lyapunov function in terms of the error quantities as 
\begin{equation} \label{eq_38}
\begin{array}{lll}
V(t)=\frac{1}{2}{\bf e}^{T}{\bf e}.
\end{array}
\end{equation}

Then using Eq.\ \eqref{eq_37}, we have 

\begin{equation} \label{eq_39}
\begin{array}{lll}
\dot{V}(t)={\bf e}^{T}\dot{\bf e}\\[10pt]
={\bf e}^{T}\bigoplus\limits_{i=1}^{N}\big[f({\bf x}_{1,i})-f({\bf x}_{1,i}-{\bf e}_i)\big]-k_A{\bf e}^{T}[\mathscr{\tilde{L}}\otimes G]{\bf e}\\[10pt]+2k_R {\bf e}^{T}[I_N\otimes H]{\bf e}\\[5pt]
=\bigoplus\limits_{i=1}^{N}{\bf e}_i^{T}\big[f({\bf x}_{1,i})-f({\bf x}_{1,i}-{\bf e}_i)\big]-k_A{\bf e}^{T}[\mathscr{\tilde{L}}\otimes G]{\bf e}\\[10pt]+2k_R {\bf e}^{T}[I_N\otimes H]{\bf e}.
\end{array}
\end{equation}

To further proceed, we have to utilize the following boundedness of the quadratic form ${\bf x}^{T}D{\bf x}$, where ${\bf x}^{T}$ denotes the transpose of ${\bf x}$ . Now if $D$ is a real symmetric matrix of order $N$, then for all ${\bf x}\in\mathbb{R}^N$
\begin{equation}\label{eq_40}
\lambda_{\min}[D]{\bf x}^{T}{\bf x}\le{\bf x}^{T}D{\bf x}\le\lambda_{\max}[D]{\bf x}^{T}{\bf x},
\end{equation}
where $\lambda_{\min}[D]$ and $\lambda_{\max}[D]$ are the minimum and maximum eigenvalues of $D$, respectively.

\par These inequalities \eqref{eq_34} and \eqref{eq_40} help to convert Eq.\ \eqref{eq_39} as follows

\begin{equation} \label{eq_41}
\begin{array}{lll}
\dot{V}(t) \leq \Big[M-k_A\lambda_{\min}[\mathscr{\tilde{L}}\otimes G]+2k_R\lambda_{\max}[I_N\otimes H]\Big]{\bf e}^{T}{\bf e}.
\end{array}
\end{equation}

Now, as per our assumption, $G$ is a positive semi-definite matrix. Therefore all the eigenvalues of $G$ are non-negative. Also, the minimum eigenvalue of $\mathscr{\tilde{L}}$ is zero. Thus, we have

\begin{equation} \label{eq_42}
\begin{array}{lll}
\lambda_{\min}[\mathscr{\tilde{L}}\otimes G]=\lambda_{\min}[\mathscr{\tilde{L}}]\lambda_{\min}[G]=0.
\end{array}
\end{equation}

Also, all the eigenvalues of $I_N$ is $1$. Thus we have

\begin{equation} \label{eq_43}
\begin{array}{lll}
\lambda_{\max}[I_N\otimes H]=\lambda_{\max}[H].
\end{array}
\end{equation}

Hence, Eq.\ \eqref{eq_41} reduces to

\begin{equation} \label{eq_44}
\begin{array}{lll}
\dot{V}(t) \leq \Big[M+2k_R\lambda_{\max}[H]\Big]{\bf e}^{T}{\bf e}.
\end{array}
\end{equation} 

Since as per assumption, $\lambda_{\max}[H]>0$. Thus, we know $\dot{V}(t)<0$ yields the required global stability condition. Hence, Eq.\ \eqref{eq_44} provides 

\begin{equation} \label{eq_45}
\begin{array}{lll}
k_R<-\dfrac{M}{2\lambda_{\max}[H]}.
\end{array}
\end{equation}

Thus, whenever we choose an interlayer coupling strength $k_R$ less than $\dfrac{-M}{2\lambda_{\max}[H]}$, the global stability of the interlayer antisynchronization state is assured irrespective of the chosen initial conditions (except for a set of measure zero) if our earlier mentioned assumptions hold. Note that the derived interlayer coupling strength is not optimized in the sense that it may be possible to calculate a better (optimal) interlayer coupling strength to achieve such an interlayer antisynchronization state by introducing more higher-order error terms in the function \eqref{eq_38}. One more noticeable thing from the relation \eqref{eq_45} is one needs negative interlayer coupling strength to establish the convergence of each oscillator in one layer anti synchronously to its counterpart oscillator on the other layer, irrespective of their initial conditions except for a set of measure zero. As per our specific choice of the interlayer coupling matrix $H=diag(1,1,1)$, we have $\lambda_{\max}[H]=1$. Thus, the required interlayer coupling strength reduces to

\begin{equation} \label{eq_46}
\begin{array}{lll}
k_R<-\dfrac{M}{2}.
\end{array}
\end{equation}

Thus, our calculated interlayer coupling strength for the global convergence to the antisynchronization state of each replica node, irrespective of initial conditions, depends crucially on the Lipschitz constant of the isolated dynamics. Hence, one requires different coupling strengths for distinct dynamical systems. For instance, the Lipschitz constant for the SL oscillator with $\omega_i=\omega=3$ is approximately $3.0$. Similarly, the Lipschitz constant for the Thomas cyclically cylindrical oscillator with $b=0.2$ is approximately $1.0711$. Therefore, the required $k_R$ for the global convergence of the replica-wise antisynchronization trajectories irrespective of initial conditions except for a set of measure zero is

$$k_R<-1.5$$ for the SL oscillators with
the chosen system parameter as the local dynamics, and
$$k_R<-0.53555$$ for the Thomas cyclically cylindrical  oscillators with the chosen system parameter as the local dynamics on top of the multiplex.

%

{\color{black}
\section{Discussions}

\par It is noteworthy that one can map the interlayer antisynchronization problem into the interlayer synchronization by applying a suitable coordinate translation, viz.\ ${\bf x}_{2,i} \to -{\bf x}^{'}_{2,i}$. Definitely, such a transformation allows us to arrive at the standard synchronization problem; hence, we can use traditional techniques to study the phenomenon. However, by doing such a transformation, we will mathematically lose some valuable information about this phenomenon. 
Our derivations indicate that we need something extra apart from being identical for obtaining interlayer antisynchronization. Specifically, the oscillators should maintain symmetry in the form of an odd evolution function, which is necessary for achieving such a novel state. Furthermore, a few steps of calculations affirm that mathematically there are no restrictions in the interlayer coupling functions for maintaining interlayer complete synchronization in the translated coordinates. However, we successfully derive that one needs odd interlayer coupling functions to obtain interlayer antisynchronization.

\par Physically, we also lose the phenomenon of antisynchronization when we apply such a coordinate transformation, as the translated coordinates lead to the standard complete synchronization phenomenon. However, such a peculiar antisynchronization state is worthy of investigation. For instance, Christiaan Huygens observed the opposite type of oscillation of two pendulums hanging from the same base in 1665 \cite{pikovsky2001universal,blekhman1988synchronization,dilao2009antiphase,carranza2016synchronization}. So, one can also map this phenomenon with ``standard complete sync" in a translated coordinate system. But we know this is a particular type of ``standard sync". Transforming the antiphase synchronization achieved by two pendulum clocks hanging on a common base into a synchronization phenomenon may solve the same problem mathematically, however, at the cost of losing the novel feature of the physical event.

\par Studying antiphase synchronization gained immense attention among researchers after the experiment by Huygens. 
 For instance, Ref.\ \cite{berner2020birth} studies the existence and stabilization of various multi-cluster states, which may not be stable (even if it exists) in single-layer networks. References \cite{berner2019multiclusters,berner2019hierarchical} investigated the partial synchronizations in the form of clusters in adaptively coupled phase oscillators. A numerical study of antiphase synchronization in a bilayer network of repulsively and bidirectionally coupled 2D lattices of van der Pol oscillators is furnished on Ref.\ \cite{shepelev2021repulsive}. Notably, the numerical study on van der Pol oscillators by Shepelev et al.\ 
supports our analytical findings too. Since its evolution function is an odd function, thus it is possible to observe interlayer antisynchronization in such a system. Using multiplex architectures in combination with attractive intralayer and repulsive interlayer connections, the antiphase synchronization of identical dynamical systems is analytically investigated in Ref.\ \cite{chowdhury2021antiphase}. This study provides an elegant way of establishing antiphase synchronization in a multiplex network by introducing repulsive coupling through any spanning tree of a single connected layer and the interlayer links.

\par In fact, there are numerous investigations on antiphase synchronization \cite{chowdhury2020effect,shepelev2021anti,liu2006antiphase,nazhan2019antiphase,kachhvah2022first} and antisynchronization \cite{li2007anti,yuan2018finite,zhangyi2022synchronization,khan2021adaptive,mahmoud2021application,chen2019anti,gowse2018co,al2010adaptive,vaidyanathan2011anti,wu2017adaptive,9745741,bhowmick2012mixed}. However, our goal is to look at the interlayer antisynchronization of attractive-repulsively coupled amplitude oscillators in a multilayer network. And each of its layers may consist of a hubs-attracting, hubs-repelling, or unweighted network. The inclusion of diverse factors like (i) attractive-repulsive interaction, (ii) multilayer networks, (iii) amplitude oscillators, and (iv) hub-attracting, hub-repelling, and unweighted intralayer adjacency matrix, leads to a complex system, and despite the complexities of our proposed model, we can provide a few exciting outcomes of this novel form of synchronization, including 
(1) necessary conditions for the existence of intralayer synchronization and interlayer antisynchronization,
(2) calculating the amplitude of each oscillator by analytically solving $2N$-coupled ordinary differential equations, (3) impact of demultiplexing, and (4) local and global stability analysis of interlayer antisynchronization state.

}

\section{Summary and remarks}

\par The present article offers a thorough understanding of interlayer antisynchronization, a novel form of synchronization that emerges in multiplex networks with hubs-attracting, hubs-repelling, and unweighted intralayer networks. Our mathematics-inspired studies allow drawing a series of important conclusions about this unique dynamical phenomenon occurring in multiplex networks in terms of its local and global stability conditions, relation to network topology, coupling functions, and robustness under demultiplexing of the network. We have demonstrated that our analytically derived conditions for the existence and stability of such a solution agree perfectly well with numerical simulations. Further, we have derived a few necessary conditions analytically for the intralayer synchronization in multiplex networks and numerically verified it by assigning two different oscillators as the local dynamics at the top of the network's nodes. Apart from that, we successfully analytically predict the SL oscillators' amplitude during the simultaneous occurrence of interlayer antisynchronization and intralayer synchronization. Our results may serve as a starting point for unveiling the novel emergent collective dynamics in various natural systems. Although we are unaware of any immediate applications of the model studied here; however, our model may prove to be beneficial for studying the complex topological behavior of brain dynamics. References \cite{de2017multilayer,buldu2018frequency} demonstrate the usefulness of studying brain dynamics using multilayer networks, and smooth brain functioning depends crucially on the co-existence of excitatory and inhibitory neurons \cite{soriano2008development,vogels2009gating}. Examining the theoretical grounds of interlayer antisynchronization is essential in gaining some intuition about the cortical neuronal networks. We conclude with the hope that our systematic investigations with the theoretical framework may offer many possibilities for future research in generic multilayer networks, revealing far more fundamental aspects of these complex forms of synchronization.\\

\section*{Acknowledgments}

\par {\color{black}We greatly appreciate the insightful comments provided by anonymous referees that helped greatly improve the manuscript.} S.N.C. wants to convey his sincere gratitude to Md Sayeed Anwar of the Indian Statistical Institute for several valuable discussions. In fact, all authors are in debt to Md Sayeed Anwar for helpful comments on the manuscript. S.N.C. also wants to thank the Department of Science and Technology, Govt. of India, for the financial support through Grant No. NMICPS/006/MD/2020-21 during the end of this work. C.H. is supported by DST-INSPIRE
Faculty Grant No. IFA17-PH193.

%
%
%

%

\typeout{}
\bibliographystyle{apsrev4-1}
\bibliography{Jeet}

\end{document}